# Exploring Urban Air Mobility Adoption Potential in San Francisco Bay Area Region: A Systems of Systems Level Case Study on Passenger Waiting Times and Travel Efficiency

Winfrey Paul Sagayam Dennis
*School of Aeronautics & Astronatics*
*Purdue University*
West Lafayette, IN
0000-0002-7087-3224

*Abstract*—**Urban Air mobility has gained momentum with recent advancements in the electric vertical take-off and landing (eVTOL) vehicles, offering faster point-to-point air taxi services that could help relieve traffic congestion in chronically overburdened cities. The research assesses the feasibility and systems-of-systems level adoption potential of UAM operations in the San Francisco Bay Area by comparing passenger departure, waiting, travel, and arrival times across key regional nodes, including San Francisco, Oakland, San Jose, and Palo Alto airports, with conventional ground transportation. A multi-agent simulation was developed in MATLAB to evaluate the fleet operations and to model demand arrival using a Poisson process under stochastic passenger flows and turnaround constraints. Results indicate that utilizing UAM during peak demand could reduce total travel times up to eighty percent across the region. The findings of this paper highlight the critical operational factors for fleet schedule optimization. Especially how the fleet size, passengers' request volumes, and turnaround time directly influence waiting time, operating cost, and overall user acceptance.**

## I. Introduction

Urban Air Mobility (UAM) is a concept that has existed since the 1940s, but does not yet exist on a large scale or autonomously. The goal of UAM is to provide an alternative mode of transportation in urban areas to reduce traffic during times of high [1]. Helicopters are a form of UAM, but are not frequently used due to their high noise, cost, and emissions [2]. With over 250 eVTOL companies developing their vehicles currently, all offer the ability to take off and land vertically, as well as to fly horizontally at high speeds, depending on their specifications, such as electric tilt rotor. These advances are enabling the design of aircraft to be more quieter and more efficient than traditional fuel-burning aircraft. Major industries like Uber [3], Boeing [4], NASA [5] all have published white papers outlining their visions for the future of Advanced Air mobility. The goal of UAM is to provide an alternative mode of transportation in urban areas to reduce traffic during times of high congestion, reduce emissions, and improve accessibility in urban areas [6].

## II. Background & Research Motivation

### A. Operational Context

A recent market study has demonstrated the potential increase in utilization rates of A recent market study has demonstrated the potential increase in utilization rates of regional service routes with new hybrid-electric regional aircraft. [7] shows more than 50% of business travelers are more likely to take a short-haul eVTOL flight through regional airports that could replace one-to-two-hour car travel commutes with 10–20 min aerial hops.

Like any other transportation system, UAM operations can also be considered as a system of systems as it requires multiple subsystems and system components to collaborate while maintaining their operational and managerial independence. With all the necessary resources for maintaining nominal operations of the vehicles, such as recharging stations, loading and unloading zones, and maintenance shops. UAM operations can be broken down into subsystem and system levels to identify potential stakeholders and their influence on the operations. When it comes to the main stakeholders of the UAM industry, they can be divided into four main categories: users, regulatory agencies, operators, and eVTOL manufacturers [8]. Users play a main role in the UAM ecosystem. They can range from individual users to urban commuters to entities such as businesses. Regulatory agencies, such as Boeing, Wisk, Archer, the FAA, and the National Aeronautics and Space Administration (NASA). The FAA will create and enforce regulations pertaining to UAM operations, and NASA, Boeing, Archer, and Wisk will develop and maintain the technology and aircraft used within the SoS. The third group of stakeholders, the operators, may include public and private transportation operators, whose focus is on making UAM appealing to users. Finally, UAM manufacturers contribute to the UAM vehicle ecosystem by researching, manufacturing, and validating technology to achieve customer satisfaction and entry into the market.

*B. Status Quo*

The vehicles and infrastructure required to make a UAM SoS do not currently exist. The current outline of the SoS dictates requirements on autonomy, aircraft operations, supervision of operations, airspace, air traffic management, third-party service providers, and vertiports [9].

Autonomy in vehicles is still in its infancy. Vehicle autonomy exists with autopilot technologies, but these systems still require pilots for their operation. Regulations of aircraft operations and air traffic management are extremely limited for smaller, urban aircraft. Existing regulations would have to be redefined to accommodate the new vehicle type and for higher volume. New regulations will need to be created for flight routes, emissions, and passenger safety. However, landing and recharging sites, known as vertiports, have not yet been constructed. Helicopter pads exist, but a larger system of vertiports would need to be created to handle the charging of the higher volume of UAM vehicles. Finally, the expectation is that an aircraft will be an electric vertical take-off and landing aircraft (eVTOL). The technology and the regulations that will need to be imposed on such technology have yet to be considered.

*C. Barriers*

One barrier of UAM is that the UAM aircraft will operate close to homes and office spaces. According to Dr. Jaiwon Shin at NASA, "communities will not accept UAM operations if the noise level significantly exceeds background noise levels" [10]. This requires the use of electric propulsion due to the high noise of traditional aircraft propulsion. As battery use in electric vehicles has increased, battery technology has become a viable method of propulsion; however, battery energy density is a significant limitation to the range of these aircraft. Another barrier to UAM is limited and changing airspace as the buildings within large cities impose limits on airspace, and temporary infrastructure, such as cranes, can cause disruptions in flight paths. Vertiport design provides another barrier because it must be designed within the regulations of the Air Carrier Access Act (ACAA) which enforces companies to provide accommodations for passengers with disabilities.

*D. Problem Definition*

The ground congestion in the San Francisco Bay Area has been the worst in the United States. With the average downtown and freeway speeds ranging from 6 mph to 26 mph [11]. With a dense population and high inter-city traffic, the San Francisco Bay Area appears to be a promising testbed for adopting UAM. For trips under 40 miles, commuters by car travelling between places like San Francisco, Oakland, San Jose, and Palo Alto endure travel times of more than 60 to 90 minutes during peak hour. Whereas UAM companies like Archer estimate an average of 10–20 minutes for electric air taxi flights [12]. However, the successful integration of such systems depends on a combination of operational, technological, and economic factors that extend beyond the aircraft's performance.

Thus, this research is motivated by two factors: (1) To quantify the operational efficiency of the eVTOL regional Connections in the SF Bay area utilizing T-100 Market data set from the Bureau of Transportation Statistics (2) Building an optimization model that incorporates aircraft performance, charging time, turnaround constraints, and passengers' wait time to validate these metrics with real-world operations.

III. METHODOLOGY

*A. UAM System Definition*

The network, which will be modelled, consists of four major key nodes from the San Francisco Bay area region airports: San Francisco Airport (SFO), Oakland Airport (OAK), San Jose Airport (SJC), and Palo Alto Airport (PAO). Each node pair (from origin to destination) is connected through direct routes defined by great-circle distances, which are the shortest path between two points on the surface of the earth. The passenger demand between node- pair is estimated from the BTS T-100 dataset, which is then normalized to account for local commuter movements. From the statistics, the daily operations of the UAM will be from 5:00 AM to 1:00 PM, with 20 hours of daily operations, with 10 minutes allocated for turnaround and charging time between each ride, and 5 minutes for taxi/take-off/landing buffer

*B. System-of-Systems Framework*

The UAM system is modeled as an interconnected Systems of systems consisting of

- Physical systems: eVTOL aircraft, vertiports, and charging infrastructure in the airports.

- Information systems: Airspace and air traffic management & Ridesharing protocols including scheduling, routing, and demand forecasting.

- Human systems: passenger waiting time and demand generation.

  The operational objective is to minimize total **effective cost**, defined as the sum of travel cost and time value, while meeting service-level constraints (waiting time ≤10 min, seat load ≥70 %) [13].

*C. Key Parameters and Constraints*

For agent-based modeling in MATLAB, the key parameters are the capacity of individual UAVs, the system status of UAVs, corresponding locations and desired destinations of Passengers, the prices of a trip with each Company's UAVs, and the time it takes Vertiport Employees to charge a UAV. Each of these affects the rate at which the passengers are serviced by the UAM system. To determine the constraints of the UAM SoS, the constraints of the UAV must be applied. The Archer Midnight Aircraft, designed by Archer, is the most recent UAV tested in the SF Bay area and will act as the base model UAV for the proposed UAM SoS. The first constraint of the SoS will be the passenger limitation per vehicle, which has a maximum capacity of four passengers. The next limiting factor that constrains the SoS is the speed at which the UAV can fly and the distance it can travel before it needs to be recharged. The Archer aircraft has a maximum range of 60 miles, which means each UAV must be flown less than 60 miles between recharging sessions; otherwise, it risks failure during flight. The average recharge time for the Archer batteries is 10 -12 minutes, which means each UAV will be inoperable for, on average, 10 minutes every 60 miles it flies. Finally, while the vehicle can safely operate at

Identify applicable funding agency here. If none, delete this text box.

an altitude of 10,000 feet, to optimize battery usage when at cruise speed, the UAV must remain between altitudes of 500 feet and 3,000 feet. The operating height helps ensure that the vehicles can operate in cities with tall buildings, without sacrificing battery or safety [14]. Limiting the operating altitude range also ensures that the short flight paths the UAV uses will not interfere with commercial aircraft flying over the same city.

Table 1. provides the detailed parameters of the Archer Midnight Aircraft and UAM network. And, Table 2. provides the latitude and longitude of the 4 airport nodes.

TABLE 1.

| Parameter | Symbol | Value |
|---|---|---|
| Cruise Speed | $V_c$ | 150 mph |
| Max Range | $R_{max}$ | 60 miles |
| Optimal Leg | $R_{opt}$ | 20 miles |
| Turnaround/Charge | $T_{chg}$ | 10 miles |
| Passenger Capacity | $N_p$ | 4 passengers |
| Operating Cost | $C_{op}$ | $605/hr |
| Value of Time | $VoT$ | $40/hr |
| Car Cost | $C_{car}$ | $0.58/mi |

TABLE 2.

| Nodes | Latitude | Longitude |
|---|---|---|
| San Francisco | 37.6190 | -122.3750 |
| Oakland | 37.7213 | -122.2210 |
| San Jose | 37.3623 | -121.9290 |
| Palo Alto | 37.4611 | -122.1150 |

The Pairwise Haversine distances has been calculated and between each node pair it ranges between 15–45 mi, which is all within Archer's 60 mi envelope.

## IV. OVERVIEW OF THE MODELLING APPROACH

An integrated Systems-of-system methodological framework was developed to evaluate systems-of-systems level Urban Air Mobility (UAM) performance in the San Francisco Bay Area. This methodology combines geographic modelling, stochastic demand generation of passengers at OD pairs, which are estimated from the T-100 dataset, optimizing the fleet, and agent-based simulation (ABM) to quantitatively assess operational feasibility and adaptation potential. The framework also facilitates the assessment of operational efficiency, UAM service, and feasibility of Archer Midnight eVTOL aircraft's service under actual operating conditions.

The methodology employs a four-layer simulation and optimization pipeline:

*A. Spatial Network Modeling*

The pairwise distances of the Origin-destination (OD) airport nodes are computed using the Haversine great-circle equation. Ensuring accurate modeling of the Archer's eVTOL mission lengths. The four major Bay area nodes (η) utilized are:

$$\eta = \{SFO, OAK, SJC, PAO\} \quad (1)$$

Each airport is located using latitude and longitude coordinates. Since, every possible OD pair (11 miles – 30 miles) Fig.1. are within Archer midnight's 60 miles envelope, every OD pair is feasible without intermediate charging breaks.

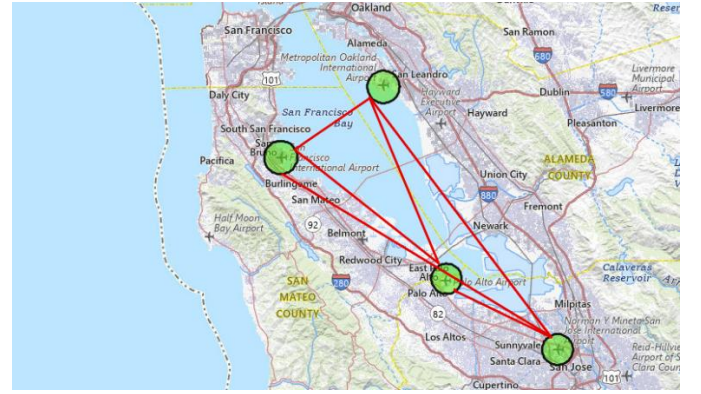

Fig.1. Flight routes considered as part of the UAM study

*B. Stochastic Passenger Demand Modeling*

To model the market demand for UAM operations, the monthly BTS T-100 Airports dataset was utilized to derive route-level passenger volumes. The origin-destination passenger counts are translated into per-minute Poisson arrival rates to simulate dynamic demand.

$$\lambda_{ij} = \frac{OD_{ij}}{30 * 20 * 60} \quad (2)$$

Where,

$$\lambda_{ij} = Passenger\ arrival\ rate$$
$$30 = days\ per\ month$$
$$20 = UAM\ operating\ hours\ per\ day$$
$$60 = minutes\ per\ hour$$

This yields arrival rates in units of passengers per minute.

The System models UAM demand as

$$N_{ij}(t) \sim Poisson\ (\lambda_{ij}.t) \quad (3)$$

$$N_{ij}(t) = No\ of\ arrivals\ in\ time\ intervals$$

Where arrivals are independent, appropriate for on-demand services.

For a simulation duration $T_{sim}$, the expected total number of arrivals during the simulation $E[N_{arrivals}]$

$$E[N_{arrivals}] = \left(\sum_{i \neq j} \lambda_{ij}\right) T_{sim} \quad (4)$$

This provides a predictive load against which fleet requirements are computed.

*C. Fleet Sizing and Optimization*

An analytical estimator is used to initialize the fleet, which is the number of aircraft required per route using flight duration.

The turnaround, charging, taxi/take-off/landing buffer constraints, and operational time windows are introduced to maintain a minimum load factor of 70–80 % while minimizing idle fleet time. Followed by simulation-driven refinement to satisfy wait time and utilization constraints.

Archer Midnight flight time between nodes is in min:

$$t_{ij}^{air} = 60 \cdot \frac{d_{ij}}{v} \quad (v = 150 \; mph) \tag{5}$$

Each mission consists of: Airborne flight, 10-minute turnaround/charging and around 5-minute taxi/take-off/landing buffer.

Thus the total cycle time is:

$$t_{ij}^{cycle} = t_{ij}^{air} + 10 + t_{ij}^{buffer} \tag{6}$$

To avoid brute-force search, an analytical model estimates the baseline fleet.

Thus, Average Cycle Time

$$\overline{t_{cycle}} = \frac{1}{|\eta|(|\eta|-1)} \sum_{i \ne j} t_{ij}^{cycle} \tag{7}$$

Cycles per Hour ($f$)

$$f = \frac{60}{\overline{t_{cycle}}} \tag{8}$$

Aircraft Passenger Capacity per Hour (with pooling) (C)

Assuming average pooling of $q = 3$ riders per flight:

$$C = f \cdot q \tag{9}$$

System Demand per Hour (D(pax/hr))

$$D = 60 \sum_{i \ne j} \lambda_{ij} \tag{10}$$

Base Fleet Estimate

$$N_{Base} = \frac{D}{C} \tag{11}$$

Robust Fleet Adjustment:

A safety factor $\alpha = 2.0$–$5.0$ compensates for demand clustering, repositioning, and imbalance:

$$N = \lceil \alpha \cdot N_{base} \rceil. \tag{12}$$

*D. Agent-Based UAM Operations Simulation*

A fully dynamic agent-based model has been utilized to simulate rider–vehicle interactions, pooling behavior, and time-based queuing at 1-minute resolution between OD pair. Each rider k is defined by OD pair, arrival time, assigned vehicle, wait time, and drop-off time. Riders enter the system via Poisson arrivals from the demand matrix. Each eVTOL vehicle has location, occupancy up to 4 passengers, state which can be idle, flying, charging, buffer/ repositioning records their time usage.

And, at each time step, the dispatch logic will identify waiting riders and available vehicles, based on which will it assign the nearest vehicle that has seat capacity, and immediate departure to rider's destination. If no vehicle is available, the rider waits.

The flight execution is implemented in the simulation utilizing

$$\Delta x = \frac{1}{t_{ij}^{air}} \tag{13}$$

Once, $|x - x_{destination}| < 1$. The simulation considers the aircraft "arrived" and it: Drops off rider(s), and enters a 10-minute charging/turnaround cycle, and becomes available again after 10 minutes. If idle and no local demand exists, vehicles reposition to the nearest high λ node to rebalance supply.

*A. Performance Evaluation*

Wait times, system throughput, and aircraft utilization are computed to validate the system feasibility.

Wait time metrics $\overline{w}$ is defined as

$$\overline{w} = \frac{1}{S} \sum_{k=1}^{S} w_k, \quad w_{95} = quantile(w_k, 0.95) \tag{14}$$

Target:

$$\overline{w} \le 10 \; minutes \tag{15}$$

There are 2 utilization.

Airborne utilization:

$$U_{air} = \frac{\sum a \, t_{air}^{(a)}}{N \, T_{sim}} \tag{16}$$

Cycle utilization (air + charging + repositioning):

$$U_{cycle} = \frac{\sum a \, t_{busy}^{(a)}}{N \, T_{sim}} \tag{17}$$

Target:

$$60\% \le U_{air}(N) \le 70\% \; air \; utlization \tag{18}$$

Throughput:

$$S_{ij} = served \; riders \; from \; i \to j$$

The code evaluated based on the feasibility feedback and the process iterates to provide the baseline sizing and simulation models validate the operational feasibility

## V. RESULTS AND DISCUSSION

This section will describe the key outcomes from these case studies in terms of the performance of the proposed UAM operations model for the San Francisco Bay Area network. The modeling evaluation incorporates demand realization, fleet throughput, passenger wait-time analysis, route-level service distribution, and aircraft utilization.

Utilizing the filtered BTS T-100 monthly, the OD passenger counts w were converted into Poisson arrival rates, yielding a total system demand of:

$$\lambda_{total} = 0.516\ pax/min \qquad (19)$$

For the 1200-minute simulation, the expected number of arrivals is:

$$E[N] = 0.516 * 1200 \approx 619.7\ riders \qquad (20)$$

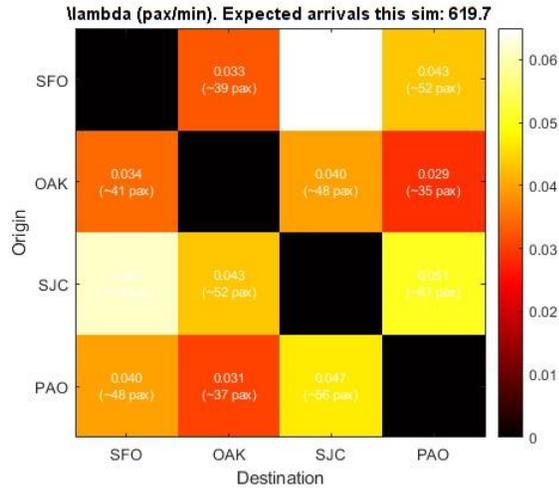

Fig.2. Demand Heatmap for Flight routes considered

591 serviced riders were generated by the simulation, which is consistent with the stochastic variation and illustrates the fleet's limited capacity under high utilization loading. The demand heatmap (Fig.2.) provides that San Jose exhibits the highest OD generation rates, followed by San Francisco while Oakland and Palo Alto provides moderate but balanced flows.

To access the system's ability to meet the service quality targets was evaluated through mean and 95th-percentile wait times.

The findings indicate:

$$Mean\ Wait\ Time: 7.47\ minutes$$
$$95th\ Percentile\ Wait\ Time: 18\ minutes$$
$$Service\ Target: \leq 10\ minutes\ (mean)$$

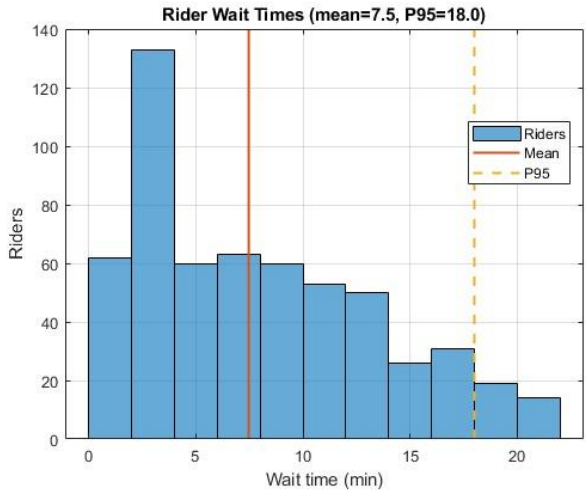

Fig.3. Rider Wait time

The wait-time histogram in Fig.3. shows that a large percentage of the riders were picked up within 5 minutes, demonstrating effective dispatch availability. Additionally, the histogram shows a right-skewed distribution that is characteristic of batch assignments and Poisson arrivals. The long-tail segment corresponds to intervals when supply is momentarily reduced due to fleet saturation and repositioning delays.

Overall, the model meets the required operational goal of less than 10-minute average wait times, validating the feasibility of maintaining acceptable quality service levels under the modelling demand.

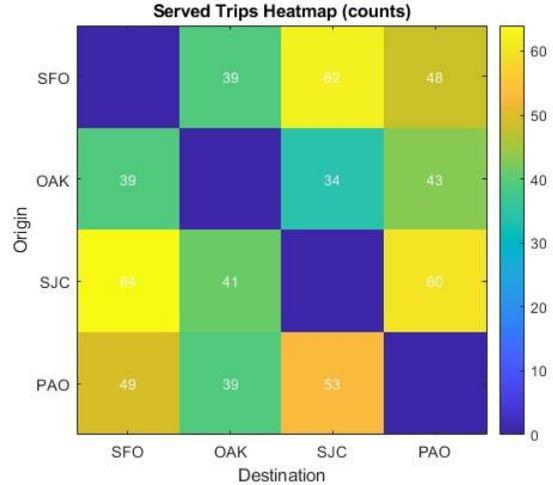

Fig.4. Served Trip Heatmap for Flight routes considered

The served-trips heatmap (Fig.4.) reveals the spatial structure of completed trips. These trips reflects two key effects: 1. San Jose dominates both inbound and outbound service since it has the greatest monthly OD flows in the filtered dataset. 2. Vehicles tend to reposition toward high λ origins, resulting in frequent flights between SJC and other airports.

Two utilization metrics are used to evaluate fleet performance. 1. The recommended Air utilization was around 60-80% but achieved 85.8%. This indicates that UAM spend most of their available time flying with minimal idle times. And the cycle utilization exceeding 90% indicates a highly saturated fleet with limited operational slack. The model therefore suggests that although 32 aircraft satisfy service quality under baseline demand, the fleet is operating at maximum sustainable load which could lead to lead to marginally overstressed utilization during peak periods. For operational robustness, a slight increase in fleet size (e.g., +10–20%) might be recommended to avoid risks congestion and long waits times during demand spikes. The system demonstrated the viability of UAM service deployment in SF Bay area by achieving complete demand coverage while retaining short waiting times.

## VI. FUTURE WORK

Future work will expand this analysis to multi-day scheduling, integrating weather and air-traffic control constraints. It would also focus on applying multi-objective optimization for cost and $CO_2$ reduction, and extend the network to include additional vertiports in the Bay area such as Berkeley, Fremont, and Santa Clara.